# gcodeml: A Grid-enabled Tool for Detecting Positive Selection in Biological Evolution

Sébastien MORETTI [a,c], Riccardo MURRI [b], Sergio MAFFIOLETTI [b], Arnold KUZNIAR [a], Briséïs CASTELLA [a], Nicolas SALAMIN [a], Marc ROBINSON-RECHAVI [a], and Heinz STOCKINGER [c,1]

[a] *Department of Ecology and Evolution, University of Lausanne and SIB Swiss Institute of Bioinformatics, Lausanne, Switzerland*
[b] *Grid Computing Competence Center (GC3), University of Zurich, Switzerland*
[c] *Vital-IT Group, SIB Swiss Institute of Bioinformatics, Lausanne, Switzerland*
[1] `Heinz.Stockinger@isb-sib.ch`

**Abstract.** One of the important questions in biological evolution is to know if certain changes along protein coding genes have contributed to the adaptation of species. This problem is known to be biologically complex and computationally very expensive. It, therefore, requires efficient Grid or cluster solutions to overcome the computational challenge. We have developed a Grid-enabled tool (*gcodeml*) that relies on the PAML (*codeml*) package to help analyse large phylogenetic datasets on both Grids and computational clusters. Although we report on results for *gcodeml*, our approach is applicable and customisable to related problems in biology or other scientific domains.

## 1 Introduction

Understanding the evolution of species is one of the main questions in biology. To address this question, it is necessary to develop mathematical models to estimate how different biological processes could have resulted in the current diversity of life. In particular, there is a strong interest to assess whether a certain feature or function could be involved in the adaptation of organisms to their environment. The force that promotes such adaptation is generally known as *positive* or *Darwinian selection* (named after the biologist Charles Darwin [2]), and the current increase of genomic data makes it possible to study the processes of adaptation at the molecular level (reviewed in [1], [4], [6]).

Selectome [14] (http://selectome.unil.ch) is a database that provides information on such positive selection events. It is provided as an on-line resource that can be easily used by life science researchers. However, the underlying computational steps are complex and require large amounts of computational resources. In brief, the computationally intensive parts are based on the phylogenetic software package called PAML [18] and in particular the *codeml* application. The focus of our work is to provide a suitable computational engine that efficiently and reliably executes thousands of *codeml* jobs on both Grid and cluster environments.

Given that multiple *codeml* jobs are independent of each other, despite sharing parts of the input data sets, the application is embarrassingly parallel and therefore suitable for cluster, cloud and Grid environments. Previous production runs have been done on the computational cluster of Vital-IT (http://www.vital-it.ch), the High Performance

Computing Center of the SIB Swiss Institute of Bioinformatics (http://www.isb-sib.ch). However, the computational needs of Selectome cannot be fully met by the Vital-IT infrastructure without seriously limiting the remaining user groups. Consequently, in a previous study we also looked into cloud solutions [10] and recently developed a solution for a Grid environment.

In the following article we present *gcodeml*, a grid-enabled version of *codeml* that runs on the ARC-based Swiss Multi-Science Computing Grid (SMSCG, http://www.smscg.ch) [17] infrastructure. Internally, the tool makes use of GC3Pie [13], a Python framework that provides the basic building blocks for high-throughput applications on both Grids (based on the Advanced Resource Connector - ARC [3]) and clusters (e.g. Sun/Oracle Grid Engine - SGE). *gcodeml* is a fault-tolerant solution that makes use of ARC's Run Time Environment (RTE) feature to execute pre-installed *codeml* applications at several sites of the SMSCG infrastructure. The underlying software system takes care of job (re)-submission and automatically corrects for most errors in case of problems with the Grid infrastructure. The latter is enabled via customised pre- and post-processing steps that follow the actual execution on remote sites. The GC3Pie framework has also been used successfully in computational chemistry [11], molecular modelling and cryptography; see http://gc3pie.googlecode.com for a list of already supported use cases.

**2 Related Work**

Addressing embarrassingly parallel applications with a cluster, cloud or Grid approach has been done for many years using various approaches and many different Grid solutions for different application domains: computational phylogeny [16], sequence search and analysis [15] as well as cryptography [9]. However, most of the existing solutions have been developed around a single use case or type of infrastructure. Extending or generalising these tools often results in the same level of complexity as developing a complete end-to-end solution. Additionally, even if Grid technologies are (slowly but progressively) reaching maturity for implementing large-scale high-throughput computational use cases, it is still required to maintain a fine-grained control of the underlying Grid services and to optimise the access to the resources. Service invocation, data transfer, fault-tolerant execution and supervision are only a few themes that have to be addressed efficiently when dealing with large-scale data analysis.

In the last few years, within the SMSCG project, we had the opportunity to identify several key high-throughput usage and scalability patterns common to many different scientific use cases. Writing end-to-end solutions for these patterns using existing tools and adopting them within the SMSCG infrastructure would have required a substantial re-engineering effort: solutions such as GANGA [12], SAGA [8] or MOTEUR [5] do not provide up-to-date interfaces to the ARC middleware, on which the SMSCG infrastructure is built. Moreover, GANGA is centered on an execution and data handling model that is more adequate to high energy physics than life science applications, making it complex to generalise.

Solutions like P-GRADE (http://portal.p-grade.hu) or WISDOM [7], despite being generic high-throughput submission engines, still require implementing the majority of the control logic (i.e., the pre- and post-processing steps). We then considered the GC3Pie framework, a Python library, to declare and supervise the execution of large application campaigns. Since GC3Pie is already available and used on the SMSCG infrastructure, it was chosen for implementing the high-level, application-specific control logic for *codeml*. Further details and background information on the GC3Pie framework will be given in Section 4.2.

## 3 Biological Background and Computational Requirements

### 3.1 The codeml Application

The *codeml* application is one of the software tools included in the PAML (Phylogenetic Analysis by Maximum Likelihood) package [18], which enables phylogenetic analyses of DNA and protein sequences using a maximum likelihood approach.

Some background: while DNA sequences are drawn from an alphabet of four different nucleotides (bases), proteins are built from 20 different amino acids. Each nucleotide triplet (e.g., AAC) is called a "codon" (cf. Figure 1) and is then translated into a unique amino acid. In brief, in order to synthesise proteins, the four-letter DNA code is translated into the 20-letter code of proteins (several codons map to the same amino acid). Therefore, the name "*codeml*", since it takes aligned codons as input, and estimates the probability of an event of positive selection in the history of the specific protein coding sequence. This is done by using the phylogenetic tree of related sequences, and then estimating the probabilities of transitions between an ancestral and descendent codon (cf. Figure 1). For instance, in an ancient species the codon TAC might have been TAT. Additionally, this change might have had a positive effect on descending species. In order to detect such changes, protein coding sequences of different species such as human, mouse or chimpanzee are compared.

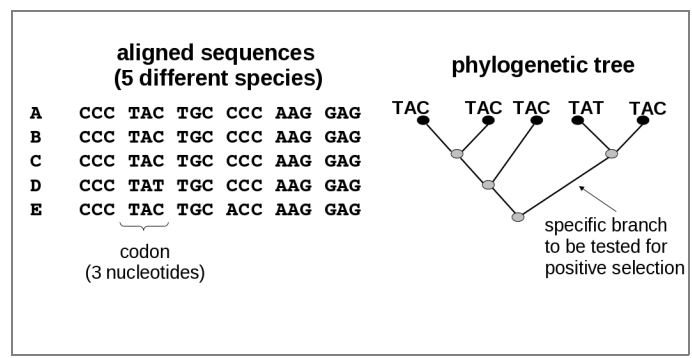

**Figure 1.** Input data for *codeml*: aligned codon sequences of five species (A-E) (left), a phylogenetic tree of the five species (right) inferred from the second column (codons) of the alignment.

In simple words, the aim of *codeml* is to detect positive selection events (resulting in improved biological functions) in the history of species. In many cases, changes within codons do not lead to positive selection events. This represents the null hypothesis (H0) that we want to test against the assumption of a positive selection event, also referred to as the alternative hypothesis (H1), using the maximum likelihood ratio test. The *codeml* program evaluates both hypotheses separately using a maximum likelihood approach.

The detailed description of the algorithm is beyond the scope of this article (cf. [18] for further details) but the important feature is that different protein coding sequences and branches can be analysed in parallel, which results in an embarrassingly parallel execution of *codeml*.

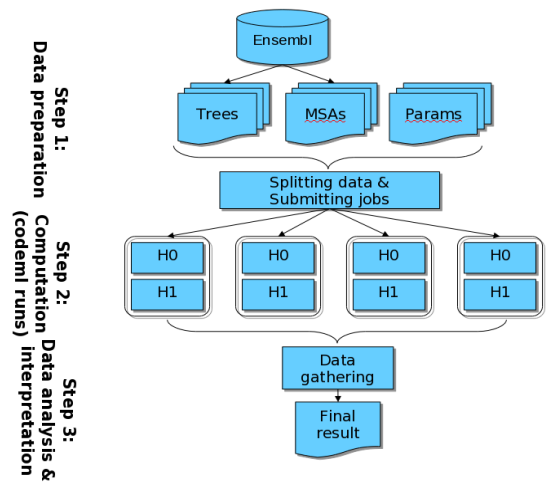

**Figure 2**. Overview of *codeml*'s dataflow using data from the Ensembl database (sequence alignments and phylogenetic trees as shown in Figure 1).

The overview of the *codeml* execution is shown in Figure 2: a single *codeml* run takes a sequence alignment file, the corresponding phylogenetic tree, as well as several parameters stored in a control file (cf. top part of Figure 2). A specific interior branch of the tree is selected to test for positive selection. Then, *codeml* is run twice, once for H0 and once for H1. Finally, data from several runs (i.e., all possible branches for each gene tree) are gathered to produce final *codeml* results, by the use of likelihood tests for each pair H0/H1, and of a Bayesian inference of selected codons.

The runtime of *codeml* usually increases with the size of the tree (i.e., the number of leaves/genes) and the length of the sequences (i.e., the number of codons). On average, a single run for a single branch of a 30-gene tree takes about 20-30 minutes to compute.

*3.2 The Selectome Database*

Given the computational costs of individual *codeml* runs, it is reasonable to pre-calculate the data for many different species and protein coding genes. In fact, this is a typical approach in bioinformatics to avoid redundant calculations done repeatedly by many groups. For positive selection, the database "Selectome" has been created as a resource for researchers in biological evolution. However, since new biological data become available with very high speed (the Ensembl database provides new releases every 2 months), Selectome needs to be regularly updated to cope with changes and new biological insights.

The results presented in Selectome are not based on simply running *codeml*: a complex workflow is involved to first prepare, then analyse, and finally to correctly display the information on a web site. For details, refer to [14] and the documentation on the Selectome web page (http://selectome.unil.ch/cgi-bin/methods.cgi). However, running *codeml* is by far the most compute-intensive part, and can only be done with large computational resources.

*3.3 Requirements*

The basic requirements for the *computational engine* of Selectome (i.e., the efficient execution of *codeml* jobs) are as follows:
- Access to large computing resources. Ensembl version 66 from February 2012 has more than 50 vertebrate species. About 20,000 phylogenetic trees are available as primary data for Selectome. This represents almost 2,000,000 jobs and 100 CPU years. The number of species sequenced increases exponentially, hence, increasing the size of the trees.
- Fault-tolerant execution. Given that tens of thousands of jobs need to be executed, a fault-tolerant execution system is required that detects potential issues and corrects them automatically: if individual jobs fail, resubmit them or retry them on other machines. The system must not lose or omit jobs (*codeml* executions) but it is free to execute them in any order.
- No human intervention to compute data: due to the scale of the problem that is directly proportional to the data size, it is necessary that the computational engine runs smoothly without human intervention, i.e., the person that launches the *codeml* runs should not need to supervise each single execution but only intervene if major technical problems occur.

**4 Architectural and Technical Details**

In the following section we describe how *gcodeml*, a client-side Grid and cluster tool, has been designed and implemented to meet the requirements. Note that *gcodeml* is not necessarily a tool that will be used by many different users but mainly by selected "production managers", such as the one that needs to provide data for the computational engine of Selectome. This is an important requirement for the design and implementation.

*4.1 Architectural Overview*

*gcodeml* is a client-side tool that uses the GC3Pie framework to manage *codeml* computational jobs, i.e., while GC3Pie provides a general job submission and execution framework (cf. Section 4.2), *gcodeml* is the application-specific code that establishes the interface between the scientist and the Grid. Similarly to the standard *codeml* program, *gcodeml* requires a set of sequence alignments, phylogenetic trees and *codeml* parameter files. Given a set of directories containing these input files, *gcodeml* creates one computational job for each pair of corresponding H0 and H1 data sets; then it submits the jobs, monitors them and retrieves output results until all jobs have been successfully executed (cf. Figure 3).

Although GC3Pie can manage jobs on a variety of batch-oriented systems, we have only used the ARC back-end to distribute jobs on the SMSCG infrastructure. In order to ensure maximum processing performance, the *codeml* application has been pre-installed at each site in SMSCG using the best available compilation options. By making use of ARC's Run Time Environment (RTE) feature, *codeml* wrapper scripts can be executed in the same way on each cluster without knowing the installation details of *codeml* at the client side. Additionally, jobs are only scheduled to sites that have the *codeml* RTE installed and functional.

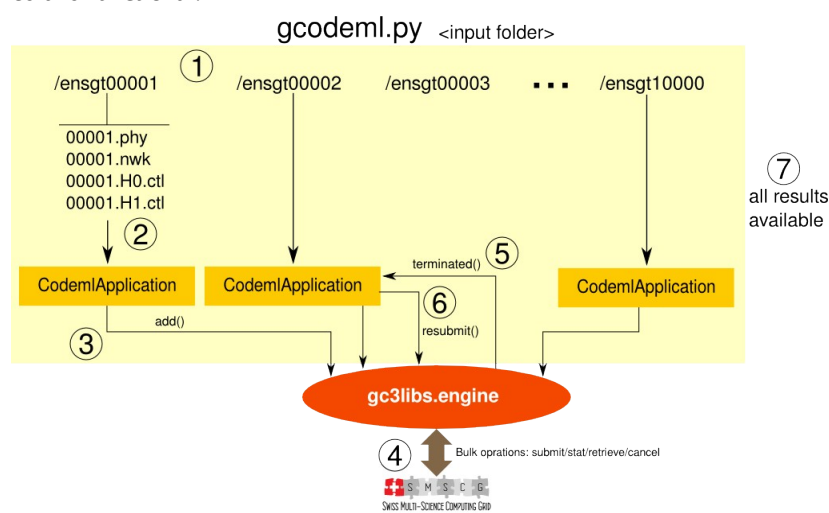

**Figure 3**. Overview of *gcodeml's workflow*: (1) input folder is provided as input argument. (2) For each valid sub-folder, an instance of *CodemlApplication* is created. (3) Each *Application* is added to a gc3libs execution engine in parallel. (4) *gc3libs.engine* executes jobs in bulk on the SMSCG infrastructure. (5) when a *CodemlApplication* is finished, its *terminated()* method is called for post-processing (6). Once all *CodemlApplication* have terminated successfully, *gcodeml* ends and results are available on specified destination (7).

*4.2 GC3Pie and gcodeml*

GC3Pie is a library of Python classes for running large job campaigns on diverse batch-oriented execution environments. At the heart of the GC3Pie model is a generic *Application* object which provides a high-level description of a computational job (including a list of input/output files, resource requirements and limits, etc.). GC3Pie translates this information into the job description format needed by the actual execution back-end selected, e.g., xRSL for ARC-based Grids, or a submission script for SGE. *Application* objects can be adapted to provide behaviour customised to a specific use case. For instance, the *CodemlApplication* object used in *gcodeml* knows how to locate the phylogenetic trees and sequence alignment files given the main *codeml* input file; in addition, it implements the post-processing hook that is run after the *codeml* job has finished and determines whether the run has been successful or not (cf. Figure 3).

GC3Pie provides composition operators that allow treating a collection of jobs as a single whole [13]: this makes it easy to implement the "embarrassingly parallel" job scheme needed by *gcodeml* by simply bundling all jobs into a single collection with no dependencies.

*4.3 Specific Features of gcodeml*

*gcodeml* is a command-line tool which scans the directory recursively for input files and creates corresponding computational jobs. Then, it manages the jobs all the way through to successful completion and retrieves output results. Processing the set of *codeml* computational jobs requires the following steps which are transparent to the *gcodeml* user except for the data organisation.

1. Input data needs to be organised such that files (sequence alignments, phylogenetic trees and parameter files for both H0 and H1) are located in the same directory. Hierarchies of directories are allowed with multiple input sets per directory in order to bundle several *codeml* runs.
2. H0 and H1 need to be computed on the same computational node in order to avoid spurious statistical deviations that might be due to different machine architectures. Therefore, corresponding H0 and H1 input files are bundled into a single *codeml* job.
3. Once the job has finished, result files are downloaded to the *gcodeml* client host or stored on a Grid Storage Element (using the GridFTP protocol). After the files have been downloaded, a post-processing step is executed: it checks that all required output files are present and parses them for specific tags to determine if *codeml* has run through the end. Sometimes, ARC might report errors although all files have been successfully processed and/or downloaded. In such cases, *gcodeml* over-writes the actual job status and corrects it.
4. If a job has failed, *gcodeml* retries it at different execution sites until it is correctly executed (i.e., the post-processing step returns an "OK" status).
5. The status of all submitted jobs is periodically monitored, and new jobs are (re)submitted until all jobs have terminated successfully.

GC3Pie takes a polling approach to managing a collection of jobs: "live" jobs (i.e., jobs that are in running or submitted/scheduled state) are checked regularly, and an appropriate action is taken depending on their status. Output from finished jobs is retrieved, and new jobs are submitted to replace the older ones. The polling interval can be set by a command-line option but, due to the ARC information system update delays, it is not useful to choose an interval shorter than 60 seconds.

In order to prevent the SMSCG infrastructure from overflow, one can specify the *gcodeml* limits for both the number of "live" jobs and the duration of a single job (which is also a safeguard against runaway *codeml* computations). The default limits are set to 50 concurrent "live" jobs of 8 hours each.

For performance reasons, *gcodeml* relies on ARC's RTE to locate clusters with pre-installed *codeml* applications. However, not each site might have *codeml* installed (this might be true for local cluster environments accessible with SGE). In such cases, the location of the *codeml* executable can be specified as a command-line argument, and *codeml* will be submitted with the actual jobs. This is also a useful feature to test new versions of *codeml*.

Given that *gcodeml* builds on ARC, standard X.509 user certificates (either long-lived ones or based on SLCS - http://www.switch.ch/grid/slcs/) are required. The GC3Pie framework (hence, *gcodeml*) takes care of proxy certificate management: if valid certificate and proxy are not available at the start of a session, users are prompted for the pass phrase; the proxy certificate is then automatically renewed when needed.

**5 Experimental Results**

A *gcodeml* client is installed on a machine at Vital-IT (within the network domain of the University of Lausanne) and is accessible for bioinformaticians who update the Selectome database. The installation uses CentOS 5.6, ARC 0.8.3, and GC3Pie 1.1 with Python 2.4, and relies on the ARC servers installed at several sites in the SMSCG infrastructure. Currently, ARC servers are installed at more than 10 sites in Switzerland, offering access to more than 7,000 CPU cores (only a subset of CPU cores is available for *gcodeml*).

In order to test both the *efficiency* and the *reliability* of our approach, we used a real dataset from the Ensembl database consisting of 1,000 gene families with 12,636 individual Grid jobs. Each job executes 2 *codeml* runs, one for H0 and one for H1. In our SMSCG infrastructure we have currently five sites that have the *codeml* RTE available and are therefore candidates for participating in the experiment. One of the sites (EPFL) consists of a Condor environment and uses a desktop Grid approach whereas all other four sites (SIB/Vital-IT (Lausanne), University of Zurich, WSL (Zurich) and HES-SO Geneva) have dedicated clusters that are accessible on a 24x7 schedule. In total, the four sites have 2,164 CPU cores but only a fraction of that is available to a single user. The exact number depends on usage and varies over time.

The aim of the experimental runs it to show that *gcodeml* fulfills the requirements of successfully executing all 12,636 *codeml* jobs in a "reasonable" amount of time (this corresponds to the CPU-intensive "Step 2" in Figure 2). It is particularly important that *gcodeml* automatically corrects for errors and delivers reliable results. Overall, we conducted the experiment three times on SMSCG (using 50, 120 and 240 "live" Grid jobs, respectively – a single Grid job can either be in running or submitted/scheduled state) as well as once on the Vital-IT cluster using 240 jobs in parallel with all data available on a parallel file system (no overhead for data transfer of input nor output). In all three Grid runs, *gcodeml* has corrected for all Grid errors and resubmitted jobs (between 17 and 118 errors were corrected) when necessary, to finally have all results correctly downloaded and stored on a client machine at Vital-IT. In the Vital-IT cluster case using LSF, no errors have occurred. Detailed performance numbers (including the number of jobs and of sites involved) are shown in Figure 4.

During one of the experiments (SMSCG 50S in Figure 4), the SLCS server (proxy certificate management server) was upgraded and, therefore, had a down-time of a few hours. However, *gcodeml* correctly recovered from this issue and demonstrated a robust submission/execution system. Due to *gcodeml*'s feature of adding new jobs and resubmitting failed jobs, gcodeml can also handle instances when a client's proxy certificate expires.

In summary, the *gcodeml* system works very reliably on several sites of the SMSCG Grid, and delivers correct results with 100% success rates. This is due to the built-in error recovery (resubmission) system. The results show that a local cluster with a parallel file system performs certainly better in terms of speed but the overall user-perceived quality of results is the same for cluster and Grid infrastructures. On both infrastructures, all test runs for detecting positive selections have been completed successfully.

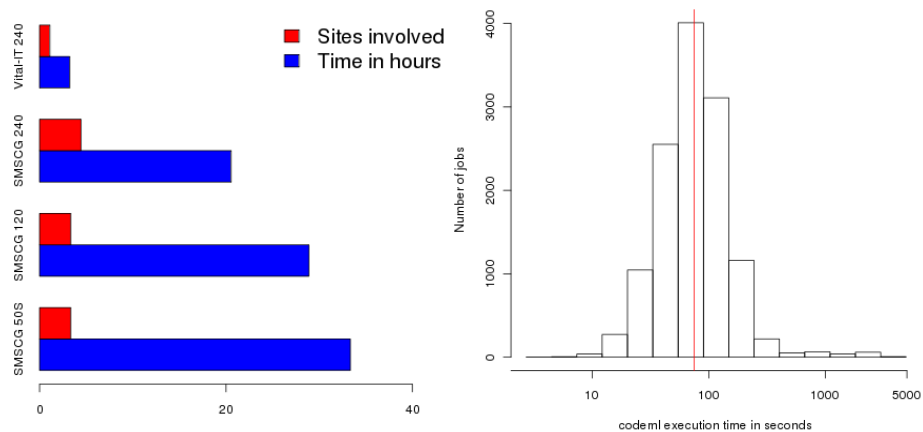

**Figure 4**. Performance of *codeml* jobs. On the left, the overall run time of all jobs is shown (including scheduling, and data transfer overhead). On the right, the individual wall times of the 12,636 *codeml* jobs are shown.

## 6 Conclusion

Our proposed system is ready to be put into production to serve an actively used biological database. The experimental runs have been successful and show that the Grid approach of *gcodeml* will allow us to recalculate Selectome for a new release based on the Ensembl database. In the future, we also aim to use the European Grid infrastructure offered by EGI either through native EMI client integration in ARC or by providing a new GC3Pie back-end for gLite Computing Elements (which are operated by many other sites in Europe, namely those participating in the WLCG Grid).

A few more technical improvements are necessary to further ease the use of the system. For instance, support for long-running jobs needs to be added based on automatic renewal of proxy certificates. Additionally, some further work is on-going with respect to handling of job status information: currently, job states are stored in the file system whereas we gain better performance with our current prototype based on a SQLite database (not yet included in *gcodeml*). This also improves access times for job status information.

Although the *codeml* algorithm is currently supporting an embarrassingly parallel approach, *codeml* does not yet make use of data-parallel features to allow for better performance of single runs. In a related project (http://www.hp2c.ch/projects/selectome/) we are currently improving both the algorithm and the implementation of the codon model used in Selectome. If the run-time of the *codeml* executable is improved, this also has a positive impact on the number of Grid and/or cluster calculations that are required to produce new versions of Selectome since many nodes are now multi-core.

**Acknowledgments**

We thank Manohar Jonnalagedda for his help in installing and configuring a first version of GC3Pie. Additionally, thanks to Volker Flegel for his support with hardware and system software. Finally, thanks to all system administrators in SMSCG for installing and supporting the *codeml* RTE. This work was supported by the projects UNIL.5 (Grid/Selectome) and SMSCG (with computational infrastructure and support) as part of the "AAA/SWITCH – e-infrastructure for e-science" programme under the leadership of SWITCH, the Swiss NREN, and has been supported by funds from the State Secretariat for Education and Research, the Federal Office for Professional Education and Technology and ETH Board. Parts of the computations were performed at the Vital-IT Center for high-performance computing of the SIB.